\documentclass{ws-ijmpcs}

\begin{document}

\markboth{Lang Shao, Fu-Wen Zhang, Zi-Gao Dai, Zhi-Ping Jin, Jing-Zhi Yan, Da-Ming Wei}
{Long and Short Gamma-Ray Bursts}

%
\catchline{}{}{}{}{}
%

\title{ON THE PUZZLE OF LONG AND SHORT GAMMA-RAY BURSTS}

\author{LANG SHAO$^{1,2,3}$, FU-WEN ZHANG$^{2,3,4}$, ZI-GAO DAI$^{5,6}$, ZHI-PING JIN$^{2,3}$, JING-ZHI YAN$^{2,3}$, DA-MING WEI$^{2,3}$ }

\address{$^1$Department of Physics, Hebei Normal University, Shijiazhang 050016, China;lang@pmo.ac.cn\\
$^2$Purple Mountain Observatory, Chinese Academy of Sciences, Nanjing 210008, China\\
$^3$Key Laboratory of Dark Matter and Space Astronomy, Chinese Academy of Sciences, Nanjing 210008, China\\
$^4$College of Science, Guilin University of Technology, Guilin 541004, China\\
$^5$School of Astronomy and Space Sciences, Nanjing University, Nanjing 210093, China\\
$^6$Key Laboratory of Modern Astronomy and Astrophysics (Nanjing University), Ministry of Education, Nanjing 210093, China}

\maketitle

\begin{history}
\received{Day Month Year}
\revised{Day Month Year}
\end{history}

\begin{abstract}
In this paper we give a brief review of our recent studies on the long and short gamma-ray bursts (GRBs) detected Swift, in an effort to understand the puzzle of classifying GRBs. We consider that it is still an appealing conjecture that both long and short GRBs are drawn from the same parent sample by observational biases. 

\end{abstract}

\ccode{PACS numbers: 98.70.Rz}

\section{Introduction}	

Before the Swift mission, it was generally accepted that there are two types of gamma-ray bursts (GRBs; long versus short) based on the bimodal distribution of their observed prompt duration\cite{kouveliotou93}. Even though this dichotomous classification scheme used to be consistent with observational and theoretical consequences, it is found controversial in the Swift era\cite{zhang11}.

Here we present the results of our latest studies of both long and short GRBs detected by Swift and address several issues that may help solve the puzzle of classifying GRBs.

\section{Distribution of duration}	

First of all, the prompt duration could not be well defined given that (1) the radiative distinction between the prompt and afterglow emission could not be well defined, let alone the fact that both the prompt and afterglow emission are very diverse in their observational properties\cite{shao10} and (2) the duration is strongly dependent on the sensitivity and bandpass of the detector and, in particular, the prompt duration of GRBs with extended emission is debatable\cite{zhang07}. A strong selection effect has been revealed regarding the distribution of the prompt duration by different detectors\cite{sakamoto08,frontera09,shao11}. The bimodal distribution is questionable and it is still an open question as to whether there are two types of GRBs.

\section{Hardness ratio}	
The hardness ratio was considered as the spectral indicator of the difference between long and short GRBs\cite{kouveliotou93}. However, it has been found that the spectral evolution of the prompt emission should be an important factor that determines the correlation between the hardness ratio and prompt duration\cite{shao11}, i.e., shorter GRBs tend to be harder and longer ones tend to be softer because of the universality of the spectral evolution of both long and short GRBs\cite{shao10,ghirlanda11}.
	
\section{Redshift distribution}	

\begin{figure}[pb]
\centerline{\psfig{file=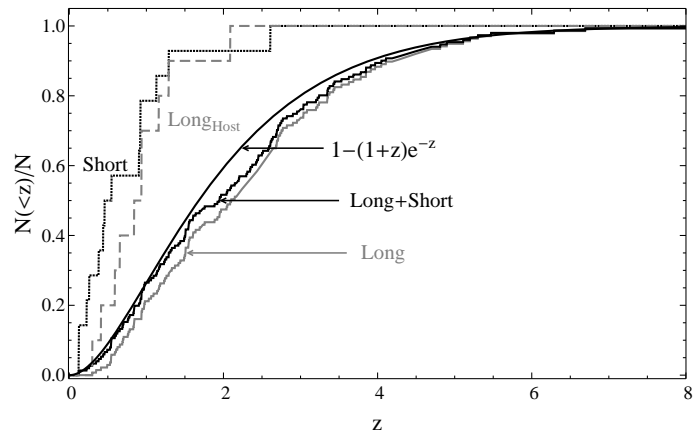,width=10cm}}
\vspace*{8pt}
\caption{Redshift distribution of 137 long and 14 short GRBs. The redshift distribution of 10 long GRBs measured from their host galaxies is shown in gray dashed steps. The solid curve is the Erlang distribution (see Ref. 7). \label{f1}}
\end{figure}

The redshift distribution has been found different for long and short GRBs. However, there are many obstacles to obtaining a reshift, making it a rare event. The distribution of all GRBs is very close to the so-called Erlang distribution 
 \begin{equation}
F(x;k,\lambda)=1-\sum_{n=0}^{k-1}e^{-\lambda x}(\lambda x)^n/n!, \label{eq:erlang}
\end{equation}
with shape paramenter $k=2$ and rate $\lambda=1$, known as the probability distribution of the waiting time until the second ``arrival'' in a one-dimensional Poisson process with a given rate (see Fig.~\ref{f1})\cite{shao11}. 

The previously considered discrepancy of redshift distribution between long and short GRBs may be caused by the observational biases. For the redshifts of 14 short GRBs in our sample, 12 of them are measured from the presumed host galaxies and only 2 of them are measured from their afterglows. For the redshifts of 137 long GRBs in our sample, only 10 of them are measured from the presumed host galaxies and the rest are measure from their afterglows. With a same measurement method, long and short GRBs appear to have a similar redshift distribution (see Fig.~\ref{f1})\cite{shao11}.

\section{Light curve}
Techniquely, short GRBs are those bursts that have a prompt duration less than 2 seconds as monitored in the Burst Alert Telescope (BAT). However, the composite X-ray light curves including both the prompt and afterglow emission suggest that most of the short GRBs might also have an intrinsically long prompt duration\cite{shao10,shao11}. If a GRB indeed have a short ($\lesssim2$~s) promt duration, the transition from the promt to the afterglow emission in 2 seconds should be recognizable on the composite light curve. Unfortunately, this case has never been clearly established for the short GRBs detected so far due to the lack of sufficient data.

\section{Conclusion}
We therefore consider that it is still an appealing conjecture that both long and short GRBs detected by Swift are drawn from the same parent sample purely by observational biases.  It is yet premature to have a dichotomous classification scheme based on the prompt duration or any other related observational properties.

\section*{Acknowledgments}

This work was supported in part by the National Natural Science Foundation of China (grants 11103083, 10973041, 10921063 and 11163003) and the National Basic Research Program of China (No. 2007CB815404). F.-W.Z. acknowledges the suport by the China Postdoctoral Science Foundation funded project (No. 20110490139), Guangxi National Science Foundation (No. 2010GXNSFB013050) and the doctoral research foundation of Guilin Univsersity of Technology.



\begin{thebibliography}{0}    

\bibitem{kouveliotou93} C. Kouveliotou, et al., {\it Astrophy. J.} {\bf 413} (1993) L101.
\bibitem{zhang11} B. Zhang, {\it Comptes Rendus Physique} {\bf 12} (2011) 206.
\bibitem{shao10} L. Shao, Y.-Z. Fan and D.-M. Wei, {\it Astrophy. J.} {\bf 719} (2010) L172.
\bibitem{zhang07} B. Zhang, et al., {\it Astrophy. J.} {\bf 655} (2007) L25.
\bibitem{sakamoto08} T. Sakamoto, et al., {\it Astrophy. J. Suppl.} {\bf 175} (2008) 179.
\bibitem{frontera09} F. Frontera, et al., {\it Astrophy. J. Suppl.} {\bf 180} (2009) 192.
\bibitem{shao11} L. Shao, Z.-G. Dai, Y.-Z. Fan, F.-W. Zhang, Z.-P. Jin and D.-M. Wei, {\it Astrophy. J.} {\bf 738} (2011) 19.
\bibitem{ghirlanda11} G. Ghirlanda, G. Ghisellini, L. Nava and D. Burlon, {\it Mon. Not. R. Astron. Soc.} {\bf 410} (2011) L47.

\end{thebibliography}
\end{document}